# Comments on: "Generalization of thermodynamics in of fractional-order derivatives and calculation of heat-transfer properties of noble gases, Journal of Thermal Analysis and Calorimetry (2018) 133:1189–1194."

I.H. Umirzakov[1]

[1] Institute of Thermophysics, 1 Lavrentev Avenue, Novosibirsk, 630090, Russia
e-mail: umirzakov@itp.nsc.ru

**Abstract**
It is shown that the equations for pressure, entropy and the isochoric heat capacity obtained by using generalization of the equilibrium thermodynamics in fractional derivatives in the paper mentioned above are approximate, the comparison of the equations with the experimental (tabulated) data for Neon and Argon made in the paper is incorrect, and the conclusions of the paper made on the basis of the comparison could be incorrect. The conditions for validity of the equations are established. It is also established that the question about a physical sense of the exponent of the derivative of a fractional order is still open.

**Keywords:** Thermodynamics, Equation of state, Entropy, Heat capacity, Fractional-order derivative

**Introduction**

The compressibility factor, entropy and isochoric heat capacity of noble gases Neon and Argon were calculated in [1] on the basis of generalization of thermodynamics in fractional derivatives using the assumption that the condition (1) (see below) is valid. We show in this paper that the equations for pressure, entropy, a compressibility factor and the isochoric heat capacity obtained in [1] are approximate; that the comparison of the equations with the experimental tabulated data for Neon and Argon is incorrect; that the conclusions of [1] made on the basis of the comparison could be incorrect; and that the use of the condition (1) alone in [1] is incorrect. The conditions for validity of the equations are established. It is also established that the question about a physical sense of the exponent of the derivative of a fractional order is still open.

**Comments**

1. The notations in this paper are the same as those in [1]. According to [1,2], Eqs. (9)-(12) [1] are valid if the following inequality takes place:

$$N^2 B / V < 1. \qquad (1)$$

The inequality (1) is valid for any values of $V$ and $N$ if the second virial coefficient $B$ is not positive. $B < 0$ if the temperature is less than the Boyle temperature $T_B$ which is defined from the condition $B(T_B) = 0$ [3]. If the temperature is greater than the Boyle temperature and, therefore, $B > 0$ [3], then we conclude from the inequality (1) that

$$BN/V < 1/N. \quad (2)$$

2. We have $\rho = 0$ from the inequality (2) in the thermodynamic limit : $N \to \infty$, $V \to \infty$ and $N/V = const$. So, Eqs. (9)-(12) [1] are not valid for macroscopic systems if the temperature is greater than the Boyle temperature.

**Table 1.** The values of maximal number of particles ($N_{max}$) which obey the condition (2).

| Neon | | Argon | |
|---|---|---|---|
| $T = 500\ K$ | $B = 6.738 \cdot 10^{-4}\ m^3/kg$ [4] | $T = 500\ K$ | $B = 1.827 \cdot 10^{-4}\ m^3/kg$ [4] |
| $\rho$, kg/m$^3$ [5] | $N_{max}$ | $\rho$, kg/m$^3$ [5] | $N_{max}$ |
| 0.49 | 3028 | 0.96 | 5701 |
| 4.84 | 306 | 9.59 | 570 |
| 9.65 | 153 | 19.15 | 285 |
| 14.42 | 102 | 28.67 | 190 |
| 19.17 | 77 | 38.16 | 143 |
| 23.88 | 62 | 47.60 | 114 |
| 28.57 | 51 | 57.00 | 96 |
| 37.84 | 39 | 75.66 | 72 |
| 47.00 | 31 | 94.12 | 58 |
| | | | |
| $T = 1000\ K$ | $B = 7.085 \cdot 10^{-4}\ m^3/kg$ [4] | $T = 1000\ K$ | $B = 5.507 \cdot 10^{-4}\ m^3/kg$ [4] |
| $\rho$, kg/m$^3$ [5] | $N_{max}$ | $\rho$, kg/m$^3$ [5] | $N_{max}$ |
| 2.42 | 583 | 4.79 | 379 |
| 12.03 | 117 | 23.73 | 76 |
| 23.87 | 59 | 46.88 | 38 |
| 46.95 | 30 | 91.48 | 19 |
| 69.30 | 20 | 133.89 | 13 |
| 90.95 | 15 | 174.23 | 10 |
| 111.96 | 12 | 212.61 | 8 |
| | | | |
| $T = 1500\ K$ | $B = 6.986 \cdot 10^{-4}\ m^3/kg$ [4] | $T = 1500\ K$ | $B = 6.058 \cdot 10^{-4}\ m^3/kg$ [4] |
| $\rho$, kg/m$^3$ [5] | $N_{max}$ | $\rho$, kg/m$^3$ [5] | $N_{max}$ |
| 1.62 | 883 | 3.20 | 515 |
| 8.05 | 177 | 15.86 | 104 |
| 16.01 | 89 | 31.43 | 52 |
| 31.66 | 45 | 61.71 | 26 |
| 46.99 | 30 | 90.89 | 18 |
| 62.00 | 23 | 119.03 | 13 |
| 76.70 | 18 | 146.18 | 11 |

Table 1 presents the maximal numbers $N_{max}$ of the particles in the system obeying the condition (2) and corresponding to the equilibrium thermodynamic states of Neon and Argon given in Tables 1 and 2 [1]. As one can see from Table 1, the maximal numbers of the particles are extremely small. They are negligible small in comparison with the Avogadro's number corresponding to the macroscopic systems. The comparison of Eqs. (9)-(12) [1] with the thermodynamic properties of Neon and Argon [4,5], which are macroscopic systems, were made in [1] for temperatures $500\,K$, $1000\,K$ and $1500\,K$, which are greater than the Boyle temperatures of Neon and Argon (the Boyle temperatures of Neon and Argon are equal to $120.32\,K$ and $407.76\,K$, respectively [6]). So, the comparison of Eqs. (9)-(12) [1] with the experimental data for Neon and Argon is incorrect.

3.   Using $B = 2\pi \int_0^\infty [1 - \exp(-u(r)/kT)] r^2 dr$, one can see that the relation $B = b - a/kT$, where $a$ and $b$ are the positive constants [7], is valid if $u(r) = +\infty$ for $r < d$, $u_{max} \equiv \max[|u(r)|, r \geq d] \neq \infty$, $-\infty < \int_d^\infty u(r) r^2 dr < 0$ and $T >> u_{max}/k$. In this case we have $b = 2\pi d^3/3$ and $a = -2\pi \int_d^\infty u(r) r^2 dr$. So, one can conclude that the relation $B = b - a/kT$ is incorrect in general case of an arbitrary pair interaction potential $u(r)$. For example, $B = b - a/kT$ is incorrect if $u(r)|_{r \to d+} = +\infty$.

4.   One can see that the values of the constants $a$ and $b$ used in Tables 1 and 2 [1] for Neon and Argon remain unknown.

5. According to [1] $U(\vec{r}_1,...,\vec{r}_N) = \sum_{1 \leq i < j \leq N} u(r_{ij}) = \sum_{i=1}^{N-1} \sum_{j=i+1}^{N} u(r_{ij})$, where $r_{ij} = |\vec{r}_i - \vec{r}_j|$ is the distance between particles with order numbers $i$ and $j$, $\vec{r}_i$, is the vector of coordinates of the $i$-th particle, so we eventually obtain from Eq. (8) [1]

$$F = -kT \ln \frac{V^N}{N! \Lambda^N} - kT \ln \left( 1 + \frac{1}{V^N} \int_V d\vec{r}_1 ... \int_V d\vec{r}_N \left( \prod_{i=1}^{N-1} \prod_{j=i+1}^{N} e^{-\frac{u(r_{ij})}{kT}} - 1 \right) \right),$$

$$F = -kT \ln \frac{V^N}{N! \Lambda^N} - kT \ln \left\{ 1 - \frac{N^2}{V} \bar{B} + \frac{1}{V^N} \int_V d\vec{r}_1 ... \int_V d\vec{r}_N \left( \prod_{i=1}^{N-1} \prod_{j=i+1}^{N} [1 - f(r_{ij})] + \sum_{i=1}^{N-1} \sum_{j=i+1}^{N} f(r_{ij}) - 1 \right) \right\}, \quad (3)$$

where $f(r_{ij}) = 1 - \exp(-u(r_{ij})/kT)$ and $\bar{B} = 1/2 \cdot \int_V f(r) d\vec{r}$. We assumed above that $N >> 1$. One can see that $\bar{B} \approx B = 2\pi \int_0^\infty f(r) r^2 dr$ if $\left| \int_V f(r) d\vec{r}/2\pi - \int_0^\infty f(r) r^2 dr \right| << \left| \int_0^\infty f(r) r^2 dr \right|$. We further assume that the last inequality is valid, so $\bar{B} = B$ in Eq. (3).

One can see from Eq. (3) using Eqs. (3)-(6) [1] that Eqs. (9)-(12) [1] obtained from

$$F = -kT \ln(V^N / N! \Lambda^N) - kT \ln(1 - N^2 B/V) \tag{4}$$

in [1,2] are incorrect.

6.   One can see that Eq. (4) presented above and used in [1,2] is valid if the inequalities (1) and

$$1 - N^2 B/V >> |\varphi|, \tag{5}$$

where $\varphi \equiv \dfrac{1}{V^N} \int\limits_V d\vec{r}_1 ... \int\limits_V d\vec{r}_N \left( \prod\limits_{i=1}^{N-1} \prod\limits_{j=i+1}^{N} [1 - f(r_{ij})] + \sum\limits_{i=1}^{N-1} \sum\limits_{j=i+1}^{N} f(r_{ij}) - 1 \right)$, are valid. The inequality (5) could be incorrect if $1 - N^2 B/V << 1$ or the density, which is proportional to $N/V$, is large.

7. It is easy to see that the integral in the right hand side of Eq. (5) [1] for $F$ from Eq. (4) exists if $B \le 0$. One can also establish that the integral in the right hand side of Eq. (6) [1] for $F$ from Eq. (4) exists for an arbitrary value of $V$ if $B \le 0$. The latter inequality is equivalent [3] to the inequality

$$T \le T_B. \qquad (6)$$

8. Using Eqs. (3) and (5) [1], one can establish from Eq. (3) that Eqs. (9) and (11) [1] may be valid if the inequalities (6) and

$$\left| \dfrac{\partial^\alpha}{\partial V^\alpha} \ln\left[ \dfrac{V^N}{N! \Lambda^N} \left(1 - \dfrac{N^2 B}{V}\right) \right] \right| >> \left| \dfrac{\partial^\alpha}{\partial V^\alpha} \left[ \ln\left(1 + \dfrac{\varphi}{1 - N^2 B/V}\right) \right] \right| \qquad (7)$$

are valid.

Using Eq. (4) and (6) [1] we conclude from Eq. (3) that Eq. (10) [1] could be valid if the inequalities (6) and

$$\left| \dfrac{\partial^\alpha}{\partial T^\alpha} \left\{ T \ln\left[ \dfrac{V^N}{N! \Lambda^N} \left(1 - \dfrac{N^2 B}{V}\right) \right] \right\} \right| >> \left| \dfrac{\partial^\alpha}{\partial T^\alpha} \left[ T \ln\left(1 + \dfrac{\varphi}{1 - N^2 B/V}\right) \right] \right| \qquad (8)$$

take place.

Using the equation $C_V = T^\alpha (\partial^\alpha S / \partial T^\alpha)_V$ for the isochoric heat capacity [2], we conclude from Eq. (3) that Eq. (12) [1] could be valid if the inequalities (6) and

$$\left| \dfrac{\partial^\alpha}{\partial T^\alpha} \dfrac{\partial^\alpha}{\partial T^\alpha} \left\{ T \ln\left[ \dfrac{V^N}{N! \Lambda^N} \left(1 - \dfrac{N^2 B}{V}\right) \right] \right\} \right| >> \left| \dfrac{\partial^\alpha}{\partial T^\alpha} \dfrac{\partial^\alpha}{\partial T^\alpha} \left[ T \ln\left(1 + \dfrac{\varphi}{1 - N^2 B/V}\right) \right] \right| \qquad (9)$$

take place.

We conclude from the consideration presented above that if the inequality (1) alone is valid, then Eqs. (9)-(12) [1] for pressure, entropy, compressibility factor and the isochoric heat capacity are incorrect. We have shown that the equations (9)-(12) [1] could be valid if the inequalities (6)-(9) are fulfilled.

We note that in the general case the validity of the inequality (5) does not mean that the inequalities (6)-(9) are valid. This can be seen from the example: $|f(x)| >> |\sin(Kx)|$ if $f(x) = x$, $K$ is an integer number, $|x| >> 1$ and $K >> 1$, but $|df(x)/dx| = 1 << K|\cos(Kx)|$ for $x = K\pi$, and $|d^2 f(x)/dx^2| = 0 << K^2 |\sin(Kx)|$ if $x = K\pi/2$ and $K$ is odd.

9. The paper [1] contains no proof of the validity of the above inequalities (6)-(9). Therefore, the statements of [1] such as "obtained results are in good agreement with the tabulated data that indicates the prospects of the proposed method for calculation of thermodynamic characteristics of a wide range of substances including both monatomic gases and complex compositions", "in the case of the equilibrium thermodynamics, the transition to fractional derivatives on the thermodynamic parameters (temperature, volume) also contributes to the theory a new parameter $\alpha$ - the rate of derivative of fractional order - with nontrivial physical sense, that implicitly takes into account the non-locality of the collision integral, thereby leading to the expansion of the area of applicability of the one-parameter family equations of state to fractal equation of state", "the results of our calculations are in satisfactory agreement with the experimentally measured data and therefore Eqs. (9)-(12) can be used for the analytical estimation of the thermal physical properties of noble gases. In addition, it is possible to extrapolate the equation of state to the extreme thermodynamic parameters, where experiments are difficult or impossible", "the measure of the fluctuations is the exponent of the derivative of a fractional order", and "as shown by our calculations for noble gases (neon and argon), the results are in satisfactory agreement with experimentally measured data. As follows from the analysis of the results, the transition to fractional derivatives with respect to time and coordinate is not a formal mathematical transition but is connected with the fundamental aspects of the physics of many-particle systems", made on the basis of the comparison, could be incorrect.

10. One can see from two equations before Eq. (8) in [1] that the Boltzmann-Gibbs distribution is used in [1] to calculate the partition function $Z$. The Boltzmann-Gibbs distribution implies that there are local and global thermal and mechanical equilibria in the system [7,8]. The use of the Boltzmann-Gibbs distribution in [1] contradicts the conclusions of [1] such as "the transition from ordinary derivatives to fractional derivatives is one of the ways of taking into account the principle of local disequilibrium, when fluctuations in thermodynamic parameters make a significant contribution to the thermodynamic process, that is, the Boltzmann–Gibbs distribution is not satisfied. In terms of its physical meaning, the transition to fractional derivatives on thermodynamic parameters is a way of taking into account the principle of local disequilibrium, when the thermodynamic process occurs under conditions of large fluctuations. That is, the transition from one equilibrium state to another does not occur through a set of equilibrium states of the system, but the contribution is made by fluctuations and non-equilibrium states that do not have time to fully relax to the equilibrium state".

11. The Boltzmann-Gibbs distribution is an equilibrium (stationary) solution of the Liouville equation [8]. The Liouville equation and Boltzmann-Gibbs distribution do not imply that the hypothesis of molecular chaos is valid [8]. The use of the Boltzmann-Gibbs distribution in [1] contradicts the statement of [1] that "To clarify the physical meaning of the derivatives of a fractional order in terms of thermodynamic parameters, we note that under conditions where the hypothesis of molecular chaos is not fulfilled, the role of fluctuations in the thermodynamic parameters becomes important. In this case, the principle of local equilibrium is violated, and the principle of local disequilibrium takes place".

12. Our detailed analysis of the text of the paper [1] shows that the paper [1] gives no proofs that "the parameter $\alpha$ - the rate of derivative of fractional order - implicitly takes into account the non-locality of the collision integral", "the measure of the fluctuations is the exponent of the derivative of a fractional order", "under conditions where the hypothesis of molecular chaos is not fulfilled, the role of fluctuations in the thermodynamic parameters becomes important" as well as "as follows from the analysis of the results, the transition to fractional derivatives with respect to time and coordinate is not a formal mathematical transition, but is connected with the fundamental aspects of the physics of many-particle systems". Therefore, we conclude that the question about the physical sense of the exponent of the derivative of a fractional order is still open.

13. As one can see from Tables 1 and 2 [1], the exponent $\alpha$ of the partial derivatives of a fractional order depends on temperature and density (volume). This dependence contradicts the assumption of [1] that $\alpha$ does not depend on temperature and volume. This assumption was used in Eqs. (5) and (6) [1] in order to obtain Eqs. (9)-(12) [1].

14. If the exponent $\alpha$ depends on temperature and volume, then it is necessary to know the dependence $\alpha(T,V)$ in order to define pressure, entropy, a compressibility factor and the isochoric heat capacity using equations (9)-(12) [1]. There is no independent definition of the exponent $\alpha(T,V)$ in [1]. Therefore, Eqs. (9)-(12) [1] could be used to describe the properties of fluid and the conclusions of [1] that "Eqs. (9)–(12) can be used for the analytical estimation of the thermal physical properties of noble gases. In addition, it is possible to extrapolate the equation of state to the extreme thermodynamic parameters, where experiments are difficult or impossible" and "…the prospects of the proposed method for calculation of thermodynamic characteristics of a wide range of substances including both monatomic gases and complex compositions" could be valid if the equation to define the dependence $\alpha(T,V)$ will be suggested.

15. It is easy to see that if the exponent $\alpha$ depends on temperature and volume then Eqs. (5) and (6) [1] must be replaced by the following definitions

$$\frac{\partial^{\alpha(T,v)} F(T,V)}{\partial V^{\alpha(T,v)}} = \frac{1}{\Gamma[1-\alpha(T,v)]} \left( \frac{\partial}{\partial V} \int_0^V \frac{F(T,v)}{(V-v)^{\alpha(T,V)}} dv \right)_T =$$
$$= \frac{1}{\Gamma[1-\alpha(T,v)]} \left( \frac{\partial}{\partial V} \int_0^V \frac{F(T,v)}{(V-v)^{\beta}} dv \right)_{T,\beta}\Bigg|_{\beta=\alpha(T,V)} - \frac{(\partial\alpha(T,V)/\partial V)_T}{\Gamma[1-\alpha(T,v)]} \cdot \int_0^V \frac{F(T,v) \cdot \ln(V-v)}{(V-v)^{\alpha(T,V)}} dv \quad , \quad (10)$$

$$\frac{\partial^{\alpha(T,v)} F(T,V)}{\partial T^{\alpha(T,v)}} = \frac{1}{\Gamma[1-\alpha(T,v)]} \left( \frac{\partial}{\partial T} \int_0^V \frac{F(t,V)}{(T-t)^{\alpha(T,V)}} dt \right)_V =$$
$$= \frac{1}{\Gamma[1-\alpha(T,v)]} \left( \frac{\partial}{\partial T} \int_0^T \frac{F(t,V)}{(T-t)^{\beta}} dt \right)_{V,\beta}\Bigg|_{\beta=\alpha(T,V)} - \frac{(\partial\alpha(T,V)/\partial T)_V}{\Gamma[1-\alpha(T,v)]} \cdot \int_0^T \frac{F(t,V) \cdot \ln(T-t)}{(T-t)^{\alpha(T,V)}} dt \quad . \quad (11)$$

Therefore, in this case it is necessary to replace Eqs. (9)-(12) [1] by those obtained by using the above two definitions of the partial derivatives. New equations will be traditional (not fractional) partial differential equations for $\alpha(T,V)$, while Eqs. (9)-(12) [1] are algebraic equations for $\alpha(T,V)$. In order to solve new equations, it is necessary to have the initial and boundary conditions for $\alpha(T,V)$. One can see from Eqs. (10) and (11) that Eqs. (9)-(12) [1] are approximately valid if the following conditions are fulfilled:

$$\left| \left( \frac{\partial}{\partial V} \int_0^V \frac{F(T,v)}{(V-v)^{\beta}} dv \right)_{T,\beta}\Bigg|_{\beta=\alpha(T,V)} \right| >> \left| \left( \frac{\partial\alpha(T,V)}{\partial V} \right)_T \cdot \int_0^V \frac{F(T,v) \cdot \ln(V-v)}{(V-v)^{\alpha(T,V)}} dv \right| ,$$

$$\left|\left(\frac{\partial}{\partial T}\int_0^T \frac{F(t,V)}{(T-t)^\beta}dt\right)_{V,\beta}\bigg|_{\beta=\alpha(T,V)}\right| >> \left|\left(\frac{\partial \alpha(T,V)}{\partial T}\right)_V \cdot \int_0^T \frac{F(t,V)\cdot \ln(T-t)}{(T-t)^{\alpha(T,V)}}dt\right|.$$

There is no proof in [1] that the above two inequalities (conditions) are fulfilled.

16. The volume $V$ is equal to the volume of the vessel in the three-dimensional space where $N$ particles are placed [7,8]. If the vessel has a fractal structure, then the use of the configurational integral $Q$, which is defined by $Q(N,V,T) \equiv \int_V d\vec{r}_1...\int_V d\vec{r}_N \exp(-U(\vec{r}_1,...,\vec{r}_N)/kT)$ and was used in [1], is incorrect because $Q(N,V,T)$ is defined by integrals over a three-dimensional volume $V$ of the vessel, which has no fractal structure, while $Q$ must be defined by integrals over a volume having a fractal structure. Therefore, Eq. (9)-(12) of [1] are not valid in this case.

**Conclusion**

So we have showed that the equations for pressure, entropy and the isochoric heat capacity obtained in [1] using a generalization of thermodynamics in fractional derivatives is approximate; the comparison of the equations with the experimental (tabulated) data for Neon and Argon [4,5] made in [1] is incorrect; the conclusions of [1] made on the basis of the comparison could be incorrect; and the use of the condition (1) alone in [1] is incorrect. The conditions for validity of the equations are established. It is also established that the question about the physical sense of the exponent of the derivative of a fractional order is still open.

It is necessary to note that the comments and conclusions presented above do not mean that the basic equations (2)-(6) [1] are incorrect. We hope that the above consideration can help to develop the generalization of thermodynamics in fractional derivatives [1,2].